\documentstyle[aps,twocolumn]{revtex}

\begin{document}
\draft
\author{Wen-Jui Huang}
\address{Department of Physics, National Changhua University of Education,Changhua
50058, Taiwan}
\title{Exact Eigenstates for Trapped Weakly Interacting Bosons in Two Dimensions }
\maketitle

\begin{abstract}
A system of $N$ two-dimensional weakly interacting bosons in a harmonic trap
is considered. When the two-particle potential is a delta function Smith and
Wilkin have analytically proved that the elementary symmetric polynomials of
particle coordinates measured from the center of mass are exact eigenstates.
In this study, we point out that their proof works equally well for an
arbitrary two-particle potential which possesses the translational and
rotational symmetries. We find that the interaction energy associated with
the eigenstate with angular momentum $L$ is equal to $aN(N-1)/2+\left(
b-a\right) NL/2$, where $a$ and $b$ are the interaction energies of two
bosons in the lowest-energy one-particle state with zero and one unit of
angular momentum, respectively. Additionally, we study briefly the case of
attractive quartic interactions. We prove rigorously that the lowest-energy
state is the one in which all angular momentum is carried by the center of
mass motion.
\end{abstract}


\smallskip The study of Bose-Einstein condensation in trapped atomic gases
has attracted a great deal of attention in the past few years \cite
{JILA,RICE,MIT,DGPS}. One of the central issues has been the possibility of
creating quantized vortices in these dilute atomic gases. Recently, vortex
states in various systems have been experimentally observed \cite
{Matthews,Madison}. Some theoretical investigations on quantized vortices
and on rotating Bose condensates have also been carried out both in the
Thomas-Fermi limit of strong interactions \cite{Rokhsar,Butts,Feder,Fetter}
and in the limit of weak interactions \cite
{Wilkin,Mottel,Bertsch,Wilkin2,Pethick,Jackson,Jackson2,Smith} between the
atoms. For the weakly interacting bosons one is naturally led to study the
model of $N$ two-dimensional bosons in a harmonic trap with weak repulsive
delta-function interactions \cite{Wilkin}. One important theoretical problem
here is to understand the properties of the system with a given total
angular momentum $L$. Numerical studies have shown that when $L>N$ the
lowest-energy state is the one where an array of singly quantized vortices
is formed \cite{Pethick,Butts}. However, due to the complexity of these
states, very few analytic results are known. In the range $0\leq L\leq N$
the structure of lowest-energy states turns out to be simpler. Numerical
computations by Bertsch and Papenbrock \cite{Bertsch} showed that the
interaction energy of the lowest-energy states has a very simple form and
decreases linearly with $L$ (the only exception being that for $L=1$). They
also noted that the wave functions for the lowest-energy states are simply
the elementary symmetric polynomials of complex coordinates relative to the
center of mass. Very recently, this remarkable formula for interaction
energy was derived analytically by Jackson and Kavoulakis \cite{Jackson2}.
Moreover, Smith and Wilkin \cite{Smith} proved that these symmetric
polynomials are indeed exact eigenstates (see Ref.\cite{Bertsch2} for an
alternative proof). In this note, we point out that the analytic proof by
Smith and Wilkin works not only for a delta-function potential but also for
an arbitrary potential which possesses translational and rotational
symmetries. In this more general case, we find that the formula for the
interaction energy still has the similar form of that for a delta-function
potential. In addition, we briefly study the case of attractive quartic
potential. We prove rigorously that, as in the cases of attractive
delta-function and harmonic potentials, the lowest-energy state is the one
where all angular momentum is carried by the center of mass motion.

Our model Hamiltonian is $\hat{H}=\hat{H}_{0}+\hat{V}$, where 
\begin{equation}
\hat{H}_{0}=\sum_{i=1}^{N}\left[ -\frac{1}{2}\nabla _{i}^{2}+\frac{1}{2}{\bf %
r}_{i}^{2}\right]
\end{equation}
is the one-particle part and 
\begin{equation}
\hat{V}=\sum_{i<j}v\left( \left| {\bf r}_{i}-{\bf r}_{j}\right| \right)
\label{potential}
\end{equation}
is the two-particle interaction. Note that the two-particle potential has
been taken to possess translational and rotational symmetries. We also
assume that the interaction term is weak and does not make the system
unstable. It is known that in the absence of interaction the one-particle
spectrum is given by 
\begin{equation}
E_{n_{r},l}=n_{r}+\left| l\right| +1,
\end{equation}
where $n_{r}$ is the radial quantum number and $l$ is the angular momentum.
For the system of $N$ non-interacting bosons, the lowest-energy states with
a given total angular momentum $L$ can be obtained by putting all bosons in
the states with $n_{r}=0$, and with $l$ of the same sign as $L$. Obviously,
there is a huge degeneracy which equals to the number of ways to distribute $%
L$ units of angular momentum among $N$ bosons.

When the two-particle interaction is added the degeneracy will, in general,
be lifted. If the interaction is weak enough, it suffices to analyze this
problem by using the first-order degenerate perturbation theory. We
therefore need to diagonalize $\hat{V}$ in the restricted Hilbert space of
lowest-energy states with angular momentum $L$. We shall do it in the second
quantized form. For a positive $L$ (as we always assume), the relevant
normalized one-particle wavefunctions for the states with $n_{r}=0$ are
given by 
\begin{equation}
\psi _{l}(r,\theta )=\frac{1}{\sqrt{l!\pi }}r^{l}e^{il\theta }e^{-r^{2}/2}=%
\frac{1}{\sqrt{l!\pi }}z^{l}e^{-\left| z\right| ^{2}/2},
\end{equation}
where $l\geq 0$ and $z=x+iy$. is the complex coordinate. We denote by $a_{l}$
and $a_{l}^{\dagger }$ the operators which annihilate and create one boson
in the one-particle state $\psi _{l}$. Our Fock space is spanned by the
basis states of the form 
\begin{equation}
\left| n_{0},n_{1,}n_{2,}...\right\rangle \equiv \Pi _{k}\frac{\left(
a_{k}^{\dagger }\right) ^{n_{k}}}{\sqrt{n_{k}!}}\left| 0\right\rangle ,
\end{equation}
where $\left| 0\right\rangle $ is the Fock vacuum and the occupation numbers 
$n_{i}^{\prime }s$ satisfy 
\begin{equation}
\sum_{k}n_{k}=N;\qquad \sum_{k}kn_{k}=L.
\end{equation}
The second quantized form of $\hat{V}$ is 
\begin{equation}
\hat{V}=\frac{1}{2}\sum_{i,j,k,l}V_{ijkl}a_{i}^{\dagger }a_{j}^{\dagger
}a_{l}a_{k},
\end{equation}
where the matrix elements are given by 
\begin{equation}
V_{ijkl}=\int \int d^{2}zd^{2}z^{\prime }\psi _{i}^{*}\left( z\right) \psi
_{j}^{*}\left( z^{\prime }\right) v(\left| z-z^{\prime }\right| )\psi
_{k}\left( z\right) \psi _{l}\left( z^{\prime }\right) .  \label{mat-element}
\end{equation}
Most of these matrix elements are actually vanishing. Indeed, the rotational
symmetry of $\hat{V}$ implies that $V_{ijkl}=0$ unless $i+j=k+l$. This, of
course, corresponds to the conservation of total angular momentum in
two-particle collisions, an important feature of a rotationally symmetric
potential. The translational symmetry of $\hat{V}$ also leads to a set of
constraints on $V_{ijkl}$. Since the potential is real, this symmetry can be
simply expressed by 
\begin{mathletters}
\begin{equation}
\sum_{i=1}^{N}\frac{\partial }{\partial z_{i}}\hat{V}=0,  \eqnum{9}
\end{equation}
where $\frac{\partial }{\partial z_{i}}=\frac{1}{2}\left( \frac{\partial }{%
\partial x_{i}}-i\frac{\partial }{\partial y_{i}}\right) $. It is easy to
check that the second quantized form of this condition reads 
\end{mathletters}
\begin{equation}
\left[ \hat{L}_{-},\hat{V}\right] =0,  \label{commu}
\end{equation}
where $\hat{L}_{-}=\sum_{k}\sqrt{k+1}a_{k}^{\dagger }a_{k+1}$. To work out
the constraints imposed by Eq.(\ref{commu}), we write 
\begin{eqnarray}
\hat{V} &=&v_{0}a_{0}^{\dagger }a_{0}^{\dagger
}a_{0}a_{0}+v_{1}a_{1}^{\dagger }a_{1}^{\dagger
}a_{1}a_{1}+v_{10}a_{1}^{\dagger }a_{0}^{\dagger }a_{1}a_{0}  \nonumber \\
&&+v_{11}a_{2}^{\dagger }a_{0}^{\dagger }a_{1}a_{1}+v_{11}^{*}a_{1}^{\dagger
}a_{1}^{\dagger }a_{2}a_{0}+\cdots .  \label{pot-form}
\end{eqnarray}
Here we have intentionally displayed only five terms since the other terms
turn out to be irrelevant for our analysis (see also Ref.\cite{Jackson2}).
Because $\hat{V}$ is hermitian, $v_{0},v_{1}$ and $v_{10}$ must be all real.
For later uses, we also note that 
\begin{equation}
v_{0}=\frac{1}{2}V_{0000};\qquad v_{1}=\frac{1}{2}V_{1111}.
\label{pot-relation}
\end{equation}
Substituting Eq.(\ref{pot-form}) into Eq.(\ref{commu}) yields 
\begin{eqnarray}
0 &=&\left[ \hat{L}_{-},\hat{V}\right]  \nonumber \\
&=&(v_{10}-2v_{0})a_{0}^{\dagger }a_{0}^{\dagger }a_{1}a_{0}+(2v_{1}-v_{10}+%
\sqrt{2}v_{11})a_{1}^{\dagger }a_{0}^{\dagger }a_{1}a_{1}  \nonumber \\
&&+\cdots .
\end{eqnarray}
Thus, the constraints, which are completely determined by the displayed
terms in Eq.(\ref{pot-form}), are 
\begin{equation}
v_{10}=2v_{0};\qquad v_{11}=\sqrt{2}(v_{0}-v_{1}).  \label{key-relation}
\end{equation}
These two constraints are direct consequences of the translational symmetry
and will play an crucial role in later analysis.

To proceed further, it is necessary to introduce the elementary symmetric
polynomials 
\begin{equation}
e_{L}=\sum_{i_{1}<i_{2}<\cdots <i_{L}}z_{i_{1}}z_{i_{2}}\cdots z_{i_{L}},
\label{symmetric}
\end{equation}
where $L\leq N$. The polynomial $e_{L}$ is related to the state $\left|
N-L,L\right\rangle $ by 
\begin{eqnarray}
&&\left\langle z_{1,}z_{2},\cdots ,z_{N}\right. \left| N-L,L\right\rangle 
\nonumber \\
&=&\left[ \frac{\left( N-L\right) !L!}{\pi ^{N}N!}\right] ^{1/2}e_{L}e^{-%
\frac{1}{2}\sum_{i=1}^{N}\left| z_{i}\right| ^{2}}\text{.}
\end{eqnarray}
Therefore, we shall use $e_{L}$ to represent the state $\left|
N-L,L\right\rangle $. In fact, it is not hard to see that there is an
one-to-one correspondence between the {\it monomial }symmetric polynomials
of degree $L$ and the basis states in our Fock space \cite{Smith}.
Consequently, each symmetric polynomial uniquely specifies a quantum state.
We also like to remark that the coordinate representation of the operator $%
\hat{L}_{+}\equiv \hat{L}_{-}^{\dagger }$ is simply $NR$, where $%
R=\sum_{i}z_{i}/N$ is the center of mass coordinate. Thus, the symmetric
polynomial $R^{k}e_{L}\left( k\geq 0\right) $ represents the state $\hat{L}%
_{+}^{k}\left| N-L,L\right\rangle $. Furthermore, since $\hat{L}_{+}$
commutes with $\hat{V}$, we have $\hat{V}\hat{L}_{+}^{k}\left|
N-L,L\right\rangle =\hat{L}_{+}^{k}\hat{V}\left| N-L,L\right\rangle $.
Equivalently, we may write it as 
\begin{equation}
\hat{V}R^{k}e_{L}=R^{k}\hat{V}e_{L}.  \label{V-on-cm}
\end{equation}
The elementary symmetric polynomials measured from the center of mass are
defined by 
\begin{equation}
\tilde{e}_{M}=\sum_{i_{1}<i_{2}<\cdots <i_{L}}\left( z_{i_{_{1}}}-R\right)
\left( z_{i_{_{2}}}-R\right) \cdots \left( z_{i_{M}}-R\right) .
\label{cm-symmetric}
\end{equation}
Note that $\tilde{e}_{1}$ is the trivial zero function. It has been proven
by Smith and Wilkin \cite{Smith} that the state represented by Eq.(\ref
{cm-symmetric}) is an exact eigenstate when $v(\left| {\bf r}\right| )$ is a
delta function. Here we follow the steps in their proof. First, we write 
\cite{Smith} 
\begin{eqnarray}
\tilde{e}_{M} &=&\sum_{L=2}^{M}\left( -1\right) ^{M-L}\frac{\left(
N-L\right) !}{(N-M)!\left( M-L\right) !}R^{M-L}e_{L}  \nonumber \\
&&+(-1)^{M-1}\frac{N!\left( M-1\right) }{(N-M)!M!}R^{M}.  \label{expansion}
\end{eqnarray}
Next, by using Eq.(\ref{pot-form}) we obtain 
\begin{eqnarray}
&&\hat{V}\left| N-L,L\right\rangle  \nonumber \\
&=&[v_{0}\left( N-L\right) (N-L-1)+v_{1}L(L-1)  \nonumber \\
&&+v_{10}(N-L)L]\left| N-L,L\right\rangle  \nonumber \\
&&+v_{11}\sqrt{L(L-1)(N-L+1)}\left| N-L+1,L-2,1\right\rangle .  \label{V-op}
\end{eqnarray}
The monomial symmetric polynomial representing the state $\left|
N-L+1,L-2,1\right\rangle $ is \cite{Smith} 
\begin{equation}
\sum_{i_{1}<i_{2}<\cdots <i_{L-2}}^{j\neq i_{1},i_{2},\cdots
i_{L-2}}z_{i_{1}}z_{i_{2}}\cdots z_{i_{L-2}}z_{j}^{2}=NRe_{L-1}-Le_{L}.
\label{A-symmetric}
\end{equation}
In terms of symmetric polynomials, Eq.(\ref{V-op}) becomes 
\begin{eqnarray}
\hat{V}e_{L} &=&[v_{0}\left( N-L\right) (N-L-1)+v_{1}L(L-1)  \nonumber \\
&&+v_{10}(N-L)L-\frac{1}{\sqrt{2}}v_{11}L(N-L+1)]e_{L}  \nonumber \\
&&+\frac{1}{\sqrt{2}}v_{11}N(N-L+1)Re_{L-1}.  \label{V-op-2}
\end{eqnarray}
It should be noted that even though Eq.(\ref{A-symmetric}) is valid only for 
$L\geq 2$ Eq.(\ref{V-op-2}) holds in the whole range $0\leq L\leq N$ if the
convention $e_{-1}=0$ is taken. Up to this point, we have not made any use
of the conditions given by Eqs.(\ref{key-relation}). To uncover the role
played by these conditions, we point out that $\tilde{e}_{M}$ is a genuine
eigenstate if Eq.(\ref{V-op-2}) has the form: 
\begin{equation}
\hat{V}e_{L}=[f(N)+Lg(N)]e_{L}-(N-L+1)g(N)Re_{L-1},  \label{V-op-3}
\end{equation}
where $f$ and $g$ are two arbitrary functions of $N$. This statement can be
easily proved by operating $\hat{V}$ on the right hand side of Eq.(\ref
{expansion}) and by using Eqs.(\ref{V-on-cm}) and (\ref{V-op-3}). We also
find that when Eq.(\ref{V-op-3}) holds the eigenvalue associated with $%
\tilde{e}_{M}$ is given by 
\begin{equation}
\varepsilon _{M,N}=f(N)+g(N)M.  \label{energy-1}
\end{equation}
It is remarkable that Eqs.(\ref{key-relation}) are precisely the required
conditions for making Eq.(\ref{V-op-2}) of the form given by Eq.(\ref{V-op-3}%
). Indeed, by substituting Eqs.(\ref{key-relation}) into Eq.(\ref{V-op-2})
we get 
\begin{eqnarray}
\hat{V}e_{L} &=&[v_{0}N(N-1)+(v_{1}-v_{0})NL]e_{L}  \nonumber \\
&&-(v_{1}-v_{0})N(N-L+1)Re_{L-1}.
\end{eqnarray}
As a result, $\tilde{e}_{M}$ is an exact eigenstate and the corresponding
eigenvalue is 
\begin{eqnarray}
\varepsilon _{M,N} &=&\left[ v_{0}N(N-1)+(v_{1}-v_{0})NM\right]  \nonumber \\
&=&\frac{1}{2}\left[ V_{0000}N(N-1)+(V_{1111}-V_{0000})NM\right] .
\label{energy-2}
\end{eqnarray}
Still we have a spectrum which varies linearly with the angular momentum $M$%
. It is worth noting that $V_{0000}$ $\left( V_{1111}\right) $ represents
the interaction energy of two bosons in the lowest-energy state with $l=0$ $%
(l=1)$. Eq.(\ref{energy-2}) represents the main result of the present work.
As a verification of this formula, we consider the delta-function potential 
\begin{equation}
v\left( {\bf r}-{\bf r}^{\prime }\right) =2\pi \eta \delta \left( {\bf r}-%
{\bf r}^{\prime }\right) ,
\end{equation}
where $\eta $ is a small dimensionless parameter. Simple calculations show 
\begin{equation}
V_{0000}=\eta ;\qquad V_{1111}=\frac{1}{2}\eta .
\end{equation}
Consequently, Eq.(\ref{energy-2}) gives 
\begin{equation}
\varepsilon _{L,N}^{delta}=\frac{\eta }{2}\left[ N(N-1)-\frac{1}{2}NL\right]
,
\end{equation}
in agreement with the results of Refs.\cite{Bertsch,Jackson2}.

Finally, we consider the case of quartic potential 
\begin{equation}
v\left( {\bf r}-{\bf r}^{\prime }\right) =\frac{\eta }{8}\left| {\bf r}-{\bf %
r}^{\prime }\right| ^{4}.
\end{equation}
When $\eta <0$ the repulsive force between two particles which are moved
away from the trap center in opposite directions would eventually be
stronger than the trapping forces acting on them. We thus expect that in
this case the harmonic trap is unable to stably confine the bosons in a
finite region of space. To avoid instability we assume that the interaction
is attractive;i.e., $\eta >0$. The purpose of studying this case is
two-fold. First, it serves as an additional example to which our results are
applied. Secondly, it provides a new case, in addition to the cases of
delta-function and harmonic potentials, where the lowest-energy states can
be analytically determined. After some straightforward algebras we get the
following expression for the two-particle interaction: 
\begin{eqnarray}
\hat{V} &=&\eta [\frac{1}{2}\hat{N}(\hat{N}-1)+\frac{7}{8}\hat{N}\hat{L}+%
\frac{1}{4}\hat{L}(\hat{L}-1)  \nonumber \\
&&+\frac{1}{8}\hat{N}\hat{J}-\frac{1}{4}(\hat{J}_{+}\hat{L}_{-}+\hat{L}_{+}%
\hat{J}_{-})+\frac{1}{8}\hat{L}_{++}\hat{L}_{--}],  \label{quart-pot}
\end{eqnarray}
where $\hat{N}=\sum_{k}a_{k}^{\dagger }a_{k}$, $\hat{L}=\sum_{k}ka_{k}^{%
\dagger }a_{k}$ and we have defined 
\begin{eqnarray}
\hat{J} &=&\sum_{k}k^{2}a_{k}^{\dagger }a_{k};\qquad  \nonumber \\
\hat{J}_{+} &=&\hat{J}_{-}^{\dagger }=\sum_{k}\left( k+2\right) \sqrt{k+1}%
a_{k+1}^{\dagger }a_{k};  \nonumber \\
\hat{L}_{++} &=&\hat{L}_{--}^{\dagger }=\sum_{k}\sqrt{(k+2)(k+1)}%
a_{k+2}^{\dagger }a_{k}.  \label{oper}
\end{eqnarray}
In this case the interaction energy associated with $\tilde{e}_{L}$ is found
to be 
\begin{equation}
\varepsilon _{L,N}^{quartic}=\frac{\eta }{2}\left[ N(N-1)+\frac{5}{2}%
NL\right] .
\end{equation}
The interaction energy increases linearly with $L$. Now $\tilde{e}_{L}$ is
no longer the lowest-energy state with angular momentum $L$. We find that,
as in the case of attractive delta-function potential \cite{Wilkin}, the
lowest-energy state is the one where all angular momentum is absorbed by the
center of mass motion. However, the proof used in Ref.\cite{Wilkin} is not
applicable here since the matrix elements $V_{ijkl}$ are not all
non-positive. We prove it by deriving a lower bound for the interaction
energy. To this end, we first note that operating $\hat{L}_{+}$ on an
eigenstate amounts to adding an unit of angular momentum to the center of
mass coordinate and the resulting state is still an eigenstate with the same
eigenvalue. Therefore, only the states, called intrinsic states and denoted
by $\left| L,N\right\rangle _{int}$, which have no center of mass
excitation, need to be considered. Such states are characterized by \cite
{Jackson2} 
\begin{equation}
\hat{L}_{-}\left| L,N\right\rangle _{int}=0.  \label{intrinsic}
\end{equation}
By using Eq.(\ref{intrinsic}), $_{int}\left\langle L,N\right| \hat{L}_{++}%
\hat{L}_{--}\left| L,N\right\rangle _{int}\geq 0$, and 
\begin{equation}
_{int}\left\langle L,N\right| \hat{J}\left| L,N\right\rangle _{int}\geq
L\;_{int}\left\langle L,N\right. \left| L,N\right\rangle _{int},\qquad
\end{equation}
we obtain 
\begin{eqnarray}
&&\frac{_{int}\left\langle L,N\right| \hat{V}\left| L,N\right\rangle _{int}}{%
_{int}\left\langle L,N\right. \left| L,N\right\rangle _{int}}  \nonumber \\
&\geq &\eta \left[ \frac{1}{2}N(N-1)+NL+\frac{1}{4}L(L-1)\right] .
\end{eqnarray}
This inequality holds for all $L\geq 0$. Since $\left| N,0\right\rangle $ is
the only eigenstate corresponding to eigenvalue $\eta N(N-1)/2$, the
lowest-energy state with a given angular momentum $L$ must be $\hat{L}%
_{+}^{L}\left| N,0\right\rangle $, the state in which all angular momentum
is in the center of mass motion. It has been argued \cite{Wilkin} (see Ref.%
\cite{Pitaevskii}, however) that this state is uncondensed and is an example
of the fragmented condensate of Nozi\`{e}res and Saint James \cite{James}.
Moreover, the presence of such an uncondensed lowest-energy state is
believed to be a general feature of the attractive interactions \cite{Wilkin}%
. Our study supplies one more example supporting this belief.

In conclusion, we have extended the work by Smith and Wilkin \cite{Smith} to
the models of $N$ bosons in a 2D harmonic trap interacting via arbitrary
rotationally and translationally symmetric potentials. We have shown that $%
\tilde{e}_{L}$ remains an exact eigenstate and the associated interaction
energy varies linearly with the angular momentum. Our analysis reveals the
importance of rotational and translational symmetries in proving this result
and makes clear the physical meaning of the coefficients appearing in the
formula for the interaction energy. We also briefly discuss the problem of
attractive quartic interactions. A lower bound for the interaction energy of
intrinsic states is derived. Based on this, we find that the lowest-energy
state for a given angular momentum $L$ is the one in which all angular
momentum is carried by the center of mass motion.

This work is supported by the National Science Council, Taiwan under Grant
No. NSC-89-2112-M-018-002.

{\it Note added}.----- A large class of interacting boson systems is
considered in a recent preprint by T. Papenbrock and G.F. Bertsch \cite
{Papen}. Our main results are in agreement with their findings.


\begin{references}
\bibitem{JILA}  M.H. Anderson, J.R. Ensher, M.R. Matthews, C.E. Wieman, and
E.A. Cornell, Science {\bf 269}, 198 (1995).

\bibitem{RICE}  C.C. Bradley, C.A. Sacket, J.J. Tollet, and R.G. Hulet,
Phys. Rev. Lett. {\bf 75}, 1687 (1995).

\bibitem{MIT}  K.B. Davis, M.-O. Mewees, M.R. Andrews, N.J. van Druten, D.S.
Durfee, D.M. Kurn, and W. Ketterle, Phys. Rev. Lett. {\bf 75}, 3969 (1995).

\bibitem{DGPS}  F. Dalfovo, S. Giogini, L.P. Pitaevskii, and S. Stringari,
Rev. Mod. Phys. {\bf 71}, 463 (1999).

\bibitem{Matthews}  M.R. Matthews, B.P. Anderson, P.C. Haljan, D.S. Hall,
C.E. Wieman, and E.A. Cornell, Phys. Rev. Lett. {\bf 83}, 2498 (1999).

\bibitem{Madison}  K.\ Madison, F. Chevy, W. Wohlleben, and J. Dalibard,
Phys. Rev. Lett. {\bf 84}, 806 (2000).

\bibitem{Rokhsar}  D.S. Rokhsar, Phys. Rev. Lett. {\bf 79}, 2164 (1997).

\bibitem{Butts}  D.A. Butts and D.S. Rokhsar, Nature {\bf 397}, 327 (1999).

\bibitem{Feder}  D.L. Feder, C.W. Clark, and B.I. Schneider, Phys. Rev.
Lett. {\bf 82}, 4956 (1999).

\bibitem{Fetter}  A.A. Svidzinsky and A.L. Fetter, Phys. Rev. Lett. {\bf 84}%
, 5919 (2000).

\bibitem{Wilkin}  N.K. Wilkin, J.M.F. Gunn, and R.A. Smith, Phys. Rev. Lett. 
{\bf 80}, 2265 (1998).

\bibitem{Mottel}  B. Mottelson, Phys. Rev. Lett. {\bf 83}, 2695 (1999).

\bibitem{Bertsch}  G.F. Bertsch and T. Papenbrock, Phys. Rev. Lett. {\bf 83}%
, 5412 (1999).

\bibitem{Wilkin2}  N.K. Wilkin, and J.M.F. Gunn, Phys. Rev. Lett. {\bf 84},
6 (2000).

\bibitem{Pethick}  G.M. Kavoulakis, B. Mottelson, and C.J. Pethick, e-print
cond-mat/0004307.

\bibitem{Jackson}  A.D. Jackson, G.M. Kavoulakis, B. Mottelson, and S.M.
Reimann, e-print cond-mat/0004309.

\bibitem{Jackson2}  A.D. Jackson, and G.M. Kavoulakis, e-print
cond-mat/0005159.

\bibitem{Smith}  R.A. Smith and N.K. Wilkin, e-print cond-mat/00005230.

\bibitem{Bertsch2}  T. Papenbrock and G.F. Bertsch, e-print cond-mat/0005480.

\bibitem{Pitaevskii}  C.J. Pethick and L.P. Pitaevskii, Phys. Rev. A {\bf 62}%
, 033609 (2000).

\bibitem{James}  P. Nozi\`{e}res and D. Saint James, J. Phys. (Paris) {\bf 43%
}, 1133 (1982).

\bibitem{Papen}  T. Papenbrock and G.F. Bertsch, e-print cond-mat/0008286.
\end{references}
\end{document}